\documentclass[12pt]{iopart}

\usepackage[dvips]{graphicx}
\begin{document}

\title[Tunable Active Brownian Motion]{Active Brownian Motion Tunable by Light}

\author{Ivo Buttinoni$^{1}$, Giovanni Volpe$^{1,2,3}$, Felix K\"{u}mmel$^{1}$, Giorgio Volpe$^{4}$ and Clemens Bechinger$^{1,2}$}
\address{$^{1}$ 2. Physikalisches Institut, Universit\"{a}t Stuttgart, Pfaffenwaldring 57, 70569 Stuttgart, Germany.}
\address{$^{2}$ Max-Planck-Institut f\"{u}r Intelligente Systeme, Heisenbergstra$\beta$e 3, 70569 Stuttgart, Germany.}
\address{$^{3}$ (present address) Department of Physics, Bilkent University, Cankaya, Ankara 06800, Turkey.}
\address{$^{4}$ ICFO-Institut de Ciencies Fotoniques, Mediterranean Technology Park, 08860, Castelldefels (Barcelona), Spain}

\pacs{82.70.Dd; 87.17.Jj;}
\submitto{\JPCM}

\begin{abstract}
Active Brownian particles are capable of taking up energy from their environment and converting it into directed motion; examples range from chemotactic cells and bacteria to artificial micro-swimmers. We have recently demonstrated that Janus particles, i.e. gold-capped colloidal spheres, suspended in a critical binary liquid mixture perform active Brownian motion when illuminated by light. In this article, we investigate in some more details their swimming mechanism leading to active Brownian motion. We show that the illumination-borne heating induces a local asymmetric demixing of the binary mixture generating a spatial chemical concentration gradient, which is responsible for the particle's self-diffusiophoretic motion. We study this effect as a function of the functionalization of the gold cap, the particle size and the illumination intensity: the functionalization determines what component of the binary mixture is preferentially adsorbed at the cap and the swimming direction (towards or away from the cap); the particle size determines the rotational diffusion and, therefore, the random reorientation of the particle; and the intensity tunes the strength of the heating and, therefore, of the motion. Finally, we harness this dependence of the swimming strength on the illumination intensity to investigate the behaviour of a micro-swimmer in a spatial light gradient, where its swimming properties are space-dependent.
\end{abstract}

\maketitle

In recent years, active Brownian motion has attracted a lot of
interest from the biological and physical communities alike
\cite{Ebbens2010}. Differently from simple Brownian particles,
whose motion is dominated by random fluctuations, active Brownian
particles feature an interplay between random fluctuations and
active swimming. While Brownian particles are at thermal
equilibrium with their environment, active Brownian particles are
capable of taking up energy and converting it into directed motion
in a process that drives them out-of-equilibrium
\cite{Erdmann2000,Hanggi2009}.

A paradigmatic example of active Brownian motion is the swimming
of bacteria such as \emph{Escherichia Coli}
\cite{Berg}. While the details may differ depending on the
specific type of bacteria, their motion can be generally described as a
sequence of roughly straight trajectories (runs) and diffusive
events (tumbles). Furthermore, bacteria can adapt their motion
to the environmental conditions, thus optimizing their survival
chances. In the presence of a nutrient gradient, for example,
bacteria are able to move towards the
higher-nutrient-concentration regions by adjusting the frequency
of their tumbles \cite{Berg1972}, employing a food searching
strategy that has been demonstrated to be  optimal when the
targets are diluted, sparse and able to regenerate
\cite{Matthaus2009}.

In recent years much effort has been devoted to realize
artificial micro-swimmers \cite{Ebbens2010}. Such  micro-swimmers
hold tremendous potential as autonomous agents to localize,
pick-up and deliver nanoscopic objects, e.g., in bioremediation,
drug-delivery and gene-therapy \cite{Weibel2005,Ford2007,vanTeeffelen2008,Downton2009,Yang2011}. Various approaches
have been proposed in order to propel such artificial micro-swimmers \cite{Ebbens2010}.
For example, micro-swimmers composed of DNA-linked magnetic colloidal beads have been powered by external rotating magnetic fields which allows the fabrication of swimmers \cite{Tierno2008,Ghosh2009,Snezhko2009}, artificial flagella
\cite{Dreyfus2005} or helical artificial tails \cite{Zhang2009}. Additionally, low-Reynolds-number swimmers have also been powered using optical
traps \cite{Leoni2008}. Although the motion of such micro-swimmers is not 
autonomous but controlled by the geometry and direction of external fields, 
the clear advantage of such approaches is that this motion can be kept alive forever.
A different approach deals with micro-swimmers capable of
autonomous motion mimicking more closely biological entities. The
main difficulty is related to the need for a constant energy (fuel) supply
that the micro-swimmers have to take up from their surrounding.
Particularly successful have been various kinds of Janus particles
driven by self-phoretic forces \cite{Golestanian2007}. Such forces
are produced by a chemical (\emph{self-diffusiophoresis}
\cite{Paxton2005,Vicario2005,Mano2005,Golestanian2005,Wang2006,Howse2007,Pantarotto2007,Palacci2010,Popescu2010}),
electrical (\emph{self-electrophoresis}) or thermal
(\emph{self-thermophoresis} \cite{Jiang2010}) gradient that the
particle is able to generate around itself. For instance, Janus
spheres or rods half coated by platinum and immersed in water
enriched by hydrogen peroxide are propelled by a catalytic
chemical reaction on the metallic surfaces
\cite{Paxton2005,Mano2005,Wang2006}; since the hydrogen peroxide
is continuously decomposed, it is necessary to adopt special
experimental configurations in order to maintain a constant
activation of the particles \cite{Palacci2010}. Self-thermophoretic micro-swimmers have also been demonstrated using
gold-capped Janus particles immersed in water \cite{Jiang2010}:
the thermophoresis occurs because the particle experience a local
temperature gradient due to the light absorbed by the gold
caps under illumination by a slightly focused laser beam. 

\begin{figure}
\centering
\includegraphics[width=12.8cm]{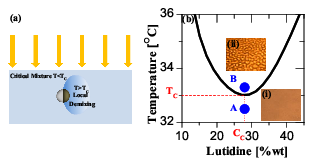}
\caption{Active Brownian micro-swimmers in a critical binary mixture.
(a) Schematic cartoon explaining the self-propulsion mechanism: a Janus particle gets illuminated, the cap is heated above $\mathrm{T_c}$ inducing a local demixing that eventually propels the particle.
(b) A schematic phase diagram for water-2,6-lutidine. The insets are bright-field microscopy pictures of the
mixed (i) and the demixed (ii) phase at the critical concentration.} \label{fig1}
\end{figure}

Very recently, we presented a novel type of micro-swimmer whose active motion relies on the local demixing of a binary mixture which is achieved by local heating of Janus particles with light (Fig.\ref{fig1}(a)) \cite{Volpe2011}. This system avoids the above mentioned fueling problem and even allows to study
active Brownian motion in patterned environments. So far, however, the origin of the propulsion mechanism has not been studied in detail. 
In this article, we investigate this swimming mechanism in more detail, showing that the local asymmetric demixing induced by the illumination-borne heating generates a local spatial chemical concentration gradient, which is eventually responsible for the particle's self-diffusiophoretic motion. We study this effect as a function of the functionalization of the gold cap, the particle size and the illumination intensity. We finally study a situation where the self-diffusiophoretic motion is spatially inhomogeneous by using an illumination gradient.

The micro-swimmers used in our experiments consist of silica spheres ($\mathrm{SiO_{2}}$,
Microparticles GmbH, Germany) with different radii ($R =
0.5,\,1,\,2.13 \,\mathrm{\mu m}$). One side of these sphere is coated by thermal evaporation with a thin adhesion layer of chromium ($2\,\mathrm{nm}$) and, subsequently,
with a thicker layer of gold ($20\,\mathrm{nm}$). In order to control which component of the mixture is preferentially adsorbed at the gold surface, we render it either hydrophilic or hydrophobic by chemical
functionalization with thiols terminated with either polar (11-Mercapto-undecanoic acid) or unpolar (1-Octadecanethiol) end groups.
These Janus
particles are then diluted in a critical binary mixture of water
and 2,6-lutidine \cite{Grattoni1993}. In figure \ref{fig1}(b) a
schematic phase diagram of the mixture is shown. At
the critical composition (0.286 mass fraction of lutidine) and below the lower
critical temperature ($\mathrm{T_c} = 307\,\mathrm{K}$) the
mixture is homogeneous, as shown by the bright-field picture in
inset (i) of figure \ref{fig1}(b). Increasing the temperature
above $\mathrm{T_c}$, the mixture separates via spinodal
decomposition as shown in inset (ii) of figure \ref{fig1}(b). The
bicontinuous wormlike structure is typical of the spinodal
decomposition and has been observed not only in binary fluids
\cite{Binder1974,Wong1981}, but also during the phase separation
of metallic alloys \cite{Gerold1978}, glass alloy systems and
polymeric blends with high molecular weight
\cite{Snyder1983,Hashimoto1986}.

The temperature of the critical suspension is kept a fraction of a degree below the critical point with a flow thermostat  (point A in figure
\ref{fig1}(b)). The whole sample ($170 \times 130\, \mathrm{\mu
m}$) is then illuminated with green light at $\lambda = 532\,\mathrm{nm}$
near the plasmonic absorption peak of gold \cite{Palik}. Due to light absorption, the gold caps are heated above $\mathrm{T_c}$; the temperature at the silica side does not change
because silica does not absorb green light.
Thus, the illumination triggers the heating of the caps and,
subsequently, the local demixing of the critical mixture. In our experiments, 
homogenous illumination is obtained by rapidly scanning a highly defocused
(spot radius $\approx 30\, \mathrm{\mu m}$) laser beam over the
entire area of the sample using an acousto-optical deflector. The
light intensity is kept lower than $\approx10\, \mathrm{\mu W/\mu m^{2}}$
in order to rule out optical forces, which are typically relevant only for $I$ at least an order of magnitude larger. 

\begin{figure}
\centering
\includegraphics[width=10cm]{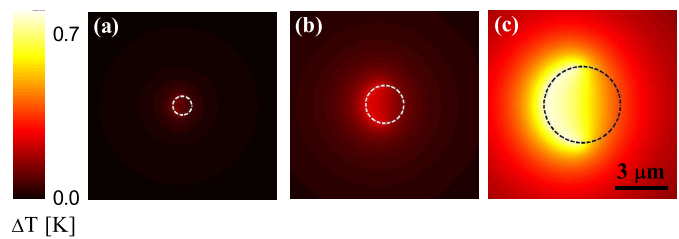}
\caption{Light-induced Heating. Temperature increase $\Delta T$ in the equatorial plane of a 
Janus particle with $R =$ (a) 0.5, (b) 1 and (c) 2 $\mathrm{\mu m}$. In all cases, the dashed line profiles the contour of the particle, the cap is on the left side and $I = 1\, \mathrm{\mu W/\mu m^{2}}$.} \label{fig2}
\end{figure}

In order to better understand the heating process induced by the illumination, we have numerically estimated the temperature increase $\Delta T$ for an incident light intensity comparable with the experimental values. Heat is generated only in the cap because of the absorption associated to the imaginary component of the dielectric constant of gold \cite{Palik} and, then, diffuses in the surrounding medium, i.e. silica sphere and water-2,6-lutidine, according to Poisson heat equation. More in detail, we followed the approach outlined in reference \cite{Baffou2009}: by applying the green dyadic method \cite{Girard2005}, we could estimate the electric field $\mathbf{E}$ inside the gold cap under uniform plane-wave illumination; then, we derived the heat power $Q$ generated at the cap, whose temperature is assumed to be uniform due to the high heat conductivity of the gold; and, finally, we calculated how $Q$ diffuses in the surrounding medium using Poisson heat equation and the heat conductivity of silica and water-2,6-lutidine. We note that $\Delta T$ is independent of the base temperature of the mixture. Figure \ref{fig2} shows the results of the simulation of $\Delta T$ in the equatorial plane of a Janus particle with $R =$ 0.5 (figure \ref{fig2}(a)), 1 (figure \ref{fig2}(b)) and 2 $\mathrm{\mu m}$ (figure \ref{fig2}(c)) illuminated with green ($532\,\mathrm{nm}$, linearly polarized) light of $I = 1\, \mathrm{\mu W/\mu m^{2}}$; brighter colors correspond to a larger $\Delta T$. The particles, whose contours are indicated by the dashed lines, are oriented so that the caps points to the left. As the particle size increases, $\Delta T$ becomes larger, because of the larger dissipation at the gold cap. More interestingly, we notice that $\Delta R$ is strongly asymmetric for all the three particle sizes and that a sharp temperature transition occurs at the edge between the capped and uncapped sides of the particle; this asymmetry is responsible for the localization of the demixing at the cap side, for the instauration of a chemical gradient across the particle and, eventually, for the micro-swimmer self-propulsion.

\begin{figure}
\centering
\includegraphics[width=12.8cm]{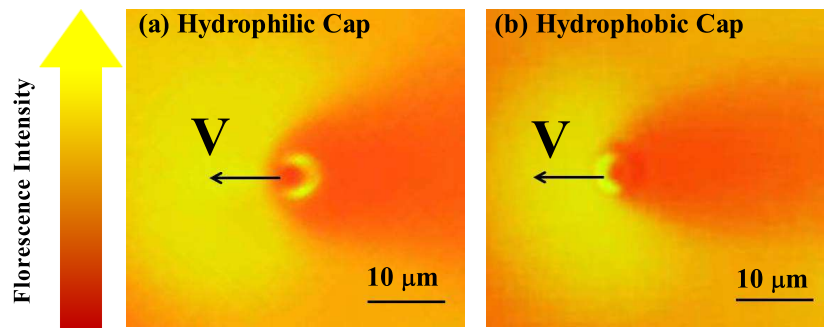}
\caption{Demixed regions around illuminated Janus particles.
(a) Distribution of the lutidine-rich phase (yellow), labelled with a hydrophobic dye (Rhodamin 6G), around a Janus particle with hydrophilic gold cap (bright half-moon shape) under illumination.
(b) Same as (a) for a Janus particle with hydrophobic gold cap. 
Since the size of the demixed regions depends on the illumination intensity, the latter is set significantly higher than in the other experiments we present ($10\,\mathrm{\mu W/\mu m^2}$ in  order to visualize better the gradient.} \label{fig3}
\end{figure}

When the temperature locally increases above $\mathrm{T_c}$, the binary mixture locally demixes and generates a demixed region around the Janus particle. In order to understand how this demixing fuels the particle's directed motion, we need to visualize its size and shape. Therefore, we added a fluorescent dye (Rhodamin 6G), which preferentially dissolves in lutidine, to the mixture and mapped the resulting fluorescence intensity. In figure \ref{fig3}, we show the fluorescence intensity encoded in colour, i.e. yellow (red) for high (low) fluorescence intensity. Figure \ref{fig3}(a)
shows the concentration of lutidine around a Janus particle with a
hydrophilic-functionalized cap. The cap is easily identified
as the bright half-moon shape at the center of the picture. Upon
illumination, the lutidine gets depleted from
the area nearby the cap (red, water-rich phase), while it
accumulates at the silica side (yellow, lutidine-rich phase), which
is hydrophobic (in comparison to the hydrophilic-functionalized
cap), thus resulting in a concentration gradient around
the particle. The particle moves within this gradient in the direction of the lutidine-rich phase as indicated by the black arrow. Figure \ref{fig3}(b) shows the lutidine
concentration around a Janus particle with a hydrophobic cap. In this
case, the lutidine-rich phase (yellow) accumulates at the cap,
while the water-rich (red) phase gets depleted and accumulates at
the silica side, being the latter more hydrophilic than
the cap. Again, the particle moves in the resulting gradient
towards the lutidine-rich phase (arrow). These measurements confirm the self-diffusiophoresis as the driving mechanism for the observed active motion. In order to rule out other possible mechanisms of self-propulsion, in particular self-thermophoresis, we repeated these measurements in pure water; here, we did not observe any directed motion under illumination intensities comparable to the ones used in our experiments.

After having investigated the driving mechanism, we studied the motion of such micro-swimmers by injecting a
diluted suspension in a thin glass cell ($\approx 100\,
\mathrm{\mu m}$ high). Owing to gravity, the particle sediment to the 
bottom of the sample cell and their motion is effectively constrained to two dimenions (for the illumination intensities considered in this work). 
We recorded movies at 7.5 frames/s and employed digital
video microscopy to track the particle trajectories at different
illumination intensities. 
When the sample is kept at a temperature far from $\mathrm{T_c}$, i.e. $\Delta T  \gg 1\, \mathrm{K}$, the local illumination-induced heating is not sufficient to cross the
spinodal line and, therefore, to induce a local demixing.
Increasing the illumination intensity, the trajectories remain Brownian with a diffusion coefficient thus given by the Stokes-Einstein formula $D_{\mathrm{SE}} = \frac{k_{B} T}{6 \pi \eta R}$,
where $\eta$ is the binary fluid viscosity \cite{Grattoni1993} and
$k_B$ is the Boltzmann constant.
When the base temperature of the sample is kept very close to $\mathrm{T_c}$, i.e. $\Delta T \approx 0.1\, \mathrm{K}$, the local heating is strong enough to cross the spinodal line and, thus, to induce a local demixing and to propel the particle. For smaller particle sizes, due to the lower heat power, higher illumination intensities are required, in agreement with the results from the simulations presented in figure \ref{fig2}.

\begin{figure}
\centering
\includegraphics[width=10cm]{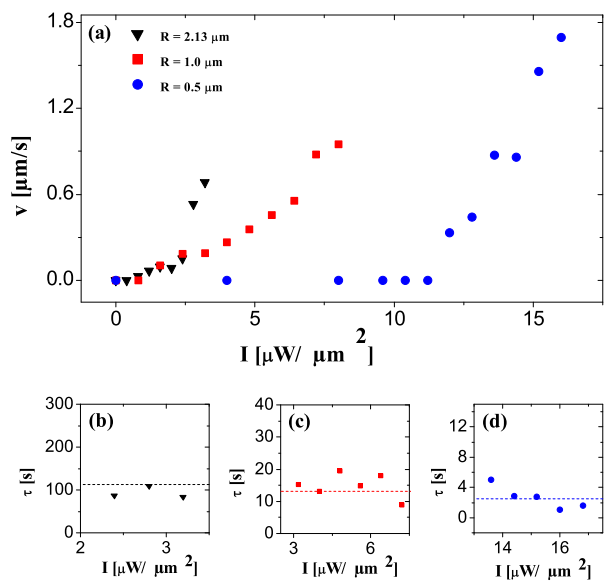}
\caption{(a) Velocity ($v$) of self-propelled Janus particles as a function of the illumination intensity. (b-d) Crossing time ($\tau$) for particles with radius $R =$ (b) 2.13, (c) 1.0 and (d) 0.5 $\mathrm{\mu m}$.} \label{fig4}
\end{figure}

The resulting motion is characterized by a crossover from ballistic motion at short times to enhanced diffusion at long times, the latter due to random changes in the swimming direction \cite{Howse2007,Palacci2010,Franke1990}. The corresponding average particle speed $v$ along the short time ballistic runs and the crossover time $\tau$ from the
ballistic to the diffusive regime can be estimated from the mean square displacement (MSD) \cite{Volpe2011}. Figure \ref{fig4} shows the resulting values. The values of $v$ as a function of $I$ are
plotted in figure \ref{fig4}(a), where the black triangles correspond to $R = 2.13\,\mathrm{\mu m}$, the red squares to $R = 1.0\,\mathrm{\mu m}$ and the blue circles to $R = 0.5\,\mathrm{\mu m}$. Clearly, active Brownian motion, i.e. $v>0$, sets in only above a certain illumination intensity threshold, which is higher for smaller particles due to their lower heat power (see also figure \ref{fig2}). This threshold can clearly be associated with the minimal heating necessary to cross the spinodal line where phase separation of the mixture occurs.

\begin{figure}
\centering
\includegraphics[width=10cm]{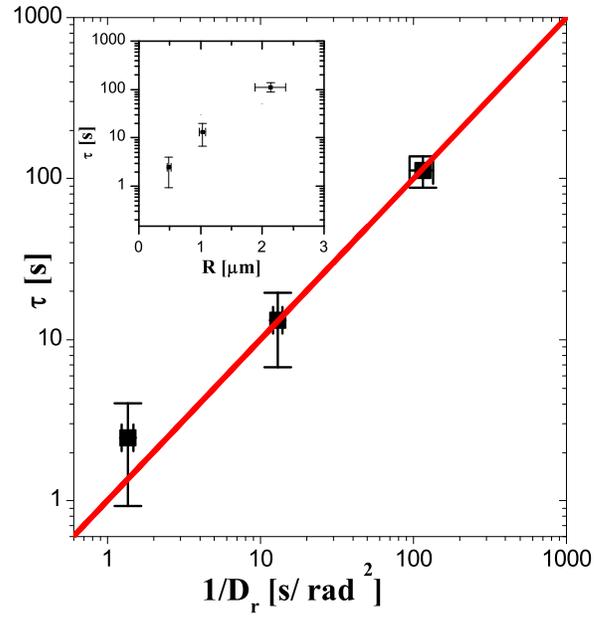}
\caption{Crossing time $\tau$ from the ballistic to the diffusive regime plotted as
a function of the rotational diffusion time $\tau_{\mathrm{r}}=1/D_{\mathrm{r}}$ over two orders of magnitude. The inset shows the $\tau$ plotted as a function of the particle radius. The error bars are obtained from the measurements at different $I$.} \label{fig5}
\end{figure}

The values of $\tau$ are shown in figure \ref{fig4}(b), \ref{fig4}(c) and \ref{fig4}(d) for $R = 2.13$, 1.0 and 0.5 $\mathrm{\mu m}$, respectively. It is evident that smaller micro-swimmers tend to change direction much more frequently. Interestingly, the values of $\tau$ for a given particle size are almost independent of the illumination intensity. These fitted values of $\tau$ agree with $\tau_{\mathrm{r}} = 1/D_{\mathrm{r}}$ within the
experimental error bars (figure \ref{fig5}), where
\begin{equation}\label{rotational}
D_{\mathrm{r}} = \frac{k_{B} T}{8 \pi \eta R^{3}}
\end{equation}
is the rotational diffusion of the particle.
This suggests that the cap reorientation is mainly due to free rotational diffusion and remains rather unaffected by the propulsion mechanism. We note that this decoupling of rotational and translational motion largely simplifies corresponding numerical simulations of this system.

\begin{figure}
\centering
\includegraphics[width=10cm]{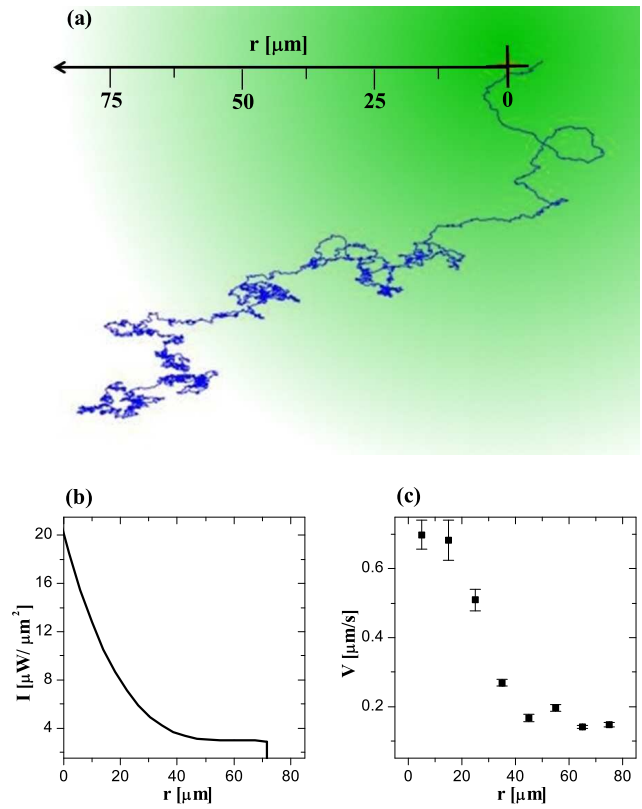}
\caption{(a) Trajectory of a self-propelled Janus particle in a radial light gradient.
(b) Intensity as a function of the radial position $r$.
(c) Average swim velocity $v$ as a function of the radial position $r$.} \label{fig6}
\end{figure}

The possibility of tuning the active Brownian motion of the
micro-swimmers can be employed in various contexts. For example, we have harnessed the dependence of the swimming strength on the 
illumination intensity to investigate the behaviour of a micro-swimmer in a 
spatial light gradient, where its swimming properties are space-dependent; these conditions resemble the situation of bacteria moving inside a chemical gradient.
Figure \ref{fig6}(a) shows the trajectory of a micro-swimmer moving inside a
radial illumination gradient; the radial dependence of the
intensity is shown in figure \ref{fig6}(b). The particle starts
from the lower left corner, where, since there is almost no light,
it undergoes standard Brownian motion. As soon as the particle
randomly moves closer to the center, it starts to perform active
Brownian motion with increasingly high $v$ as it approaches the
high-intensity center of the gradient; the radial dependence of
$v$ is plotted in figure \ref{fig6}(c). Furthermore, the
trajectories become more directed and less rough as the active
motion increases.

In conclusion, we have described a self-propulsion mechanism based on the local asymmetric demixing of a critical binary mixture around a microscopic Janus particle. The main advantage of this mechanism is that, since the required heating is very small (a fraction of a Kelvin), the active Brownian motion can be easily tuned by a very weak illumination, which permits us to avoid optical forces acting on the highly asymmetric Janus particles. Furthermore, it is possible to control the active Brownian motion of the micro-swimmers both in space and in time by employing spatial and temporal illumination patterns.

We gratefully acknowledge D. Vogt and H.-J. K\"ummerer for their help with the experiments. This work has been partially supported by the Marie Curie-Initial Training Network Comploids, funded by the European Union Seventh Framework Program (FP7).

\section*{References}

\newpage

\end{document}